\documentclass[sigconf]{acmart}
\usepackage{multirow}
\usepackage{subfigure}
\usepackage{amsmath}
\usepackage{algorithm}
\usepackage{algorithmic}

\AtBeginDocument{%
  \providecommand\BibTeX{{%
    \normalfont B\kern-0.5em{\scshape i\kern-0.25em b}\kern-0.8em\TeX}}}

\setcopyright{acmcopyright}
\copyrightyear{2021}
\acmYear{2021}
\acmDOI{10.1145/3404835.3463248}

\acmConference[SIGIR '21] {Proceedings of the 44th International ACM SIGIR Conference on Research and Development in Information Retrieval}{July 11--15, 2021}{Virtual Event, Canada}
\acmBooktitle{Proceedings of the 44th International ACM SIGIR Conference on Research and Development in Information Retrieval (SIGIR '21), July 11--15, 2021, Virtual Event, Canada}
\acmPrice{15.00}
\acmISBN{978-1-4503-8037-9/21/07}

\begin{document}
\title{EXTRA: Explanation Ranking Datasets for Explainable Recommendation}


\author{Lei Li}
\affiliation{%
	\institution{Hong Kong Baptist University}
	\city{Hong Kong}
	\country{China}
}
\email{csleili@comp.hkbu.edu.hk}

\author{Yongfeng Zhang}
\affiliation{%
	\institution{Rutgers University}
	\city{New Brunswick}
	\country{USA}
}
\email{yongfeng.zhang@rutgers.edu}

\author{Li Chen}
\affiliation{%
	\institution{Hong Kong Baptist University}
	\city{Hong Kong}
	\country{China}
}
\email{lichen@comp.hkbu.edu.hk}

\renewcommand{\shortauthors}{Li and Zhang, et al.}

\begin{abstract}
Recently, research on explainable recommender systems has drawn much attention from both academia and industry, resulting in a variety of explainable models. As a consequence, their evaluation approaches vary from model to model, which makes it quite difficult to compare the explainability of different models. To achieve a standard way of evaluating recommendation explanations, we provide three benchmark datasets for EXplanaTion RAnking (denoted as EXTRA), on which explainability can be measured by ranking-oriented metrics. Constructing such datasets, however, poses great challenges. First, user-item-explanation triplet interactions are rare in existing recommender systems, so how to find alternatives becomes a challenge. Our solution is to identify nearly identical sentences from user reviews. This idea then leads to the second challenge, i.e., how to efficiently categorize the sentences in a dataset into different groups, since it has quadratic runtime complexity to estimate the similarity between any two sentences. To mitigate this issue, we provide a more efficient method based on Locality Sensitive Hashing (LSH) that can detect near-duplicates in sub-linear time for a given query. Moreover, we make our code publicly available to allow researchers in the community to create their own datasets.
\end{abstract}



\maketitle

\section{Introduction}

Explainable Recommender Systems (XRS) \cite{SIGIR14-EFM, ACL21-PETER, FTIR20-Survey} that not only provide users with personalized recommendations but also justify why they are recommended, have become an important research topic in recent years.
Compared with other recommendation algorithms such as Collaborative Filtering (CF) \cite{WWW01-ICF, CSCW94-UCF} and Collaborative Reasoning (CR) \cite{shi2020neural, chen2021neural} which aim to tackle the information overload problem for users, XRS further improves users' satisfaction and experience \cite{SIGIR14-EFM, FTIR20-Survey, Handbook15-Explanation} by helping them to better understand the recommended items.
Actually, explanation is as important as recommendation itself, because usually there is no absolute ``right'' or ``wrong'' in terms of which item(s) to recommend.
Instead, multiple items may all be of interest to the user, and it all depends on how we explain our recommendation to users.

However, as explanations can take various forms, such as pre-defined template \cite{SIGIR14-EFM, JIIS20-CAESAR}, generated text \cite{EARS19-HSS, CIKM20-NETE, ACL21-PETER} or decision paths on knowledge graphs \cite{SIGIR19-PGPR, xian2020cafe, fu2020fairness, zhao2020leveraging}, it is sometimes difficult to evaluate the explanations produced by different methods. 

In this work, with the aim to make the standard evaluation of explainable recommendation possible, we present three benchmark datasets on which recommendation explanations can be evaluated quantitatively via standard ranking metrics, such as NDCG, Precision and Recall.
The idea of explanation ranking is inspired by information retrieval which intelligently ranks available contents (e.g., documents or images) for a given query.
In addition, this idea is also supported by our observation on the problems of existing natural language generation techniques.
In our previous work on explanation generation \cite{CIKM20-NETE}, we find that a large amount of the generated sentences are commonly seen sentences in the training data, e.g., ``\textit{the food is good}'' as an explanation for a recommended restaurant.
This means that the generation models are fitting the given samples rather than creating new sentences.
Furthermore, even strong language models such as Transformer \cite{NIPS17-Transformer} trained on a large text corpus may generate contents that deviate from facts, e.g., ``\textit{four-horned unicorn}'' \cite{19-GPT2}.

\begin{figure}
	\centering
	\includegraphics[scale=0.45]{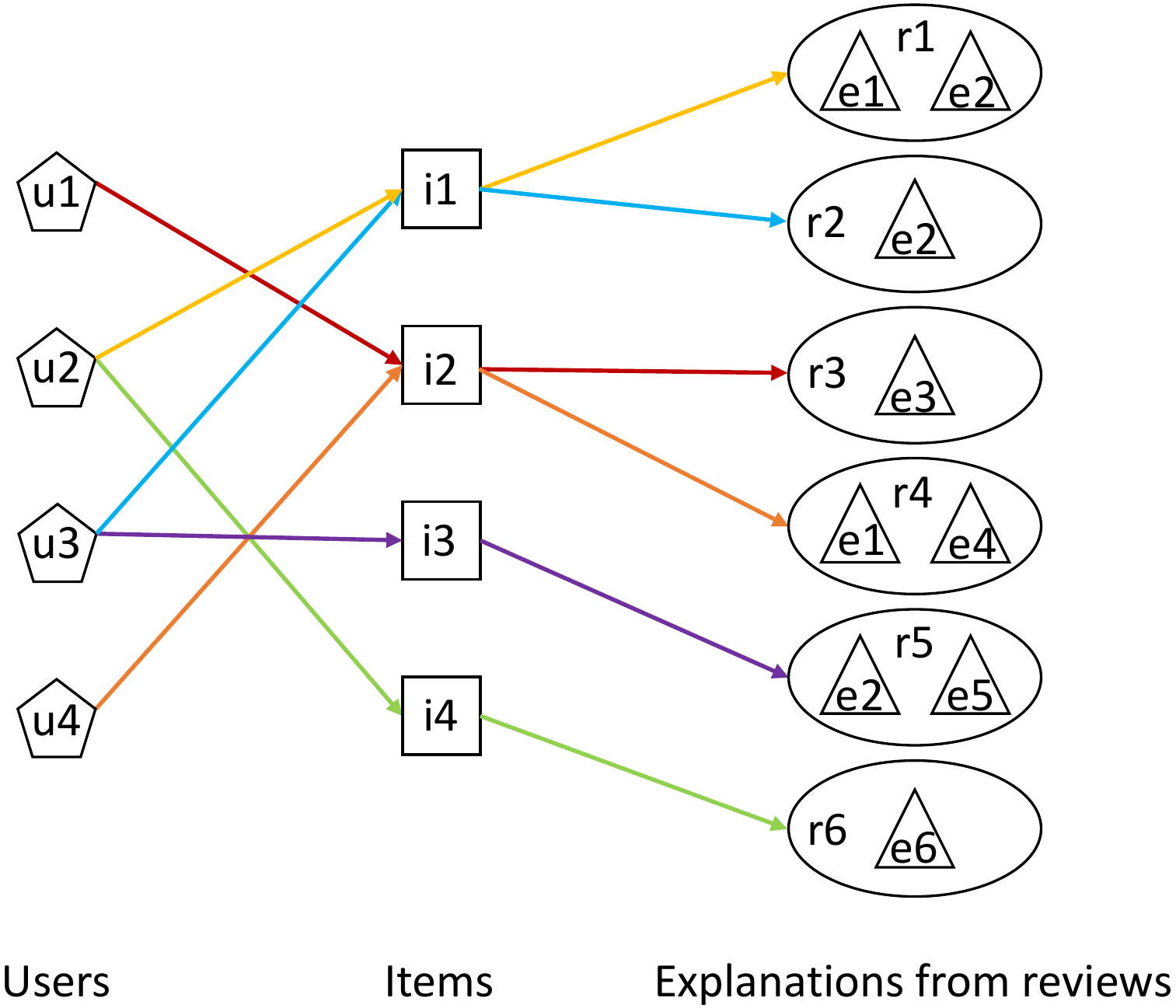}
	\caption{User-item-review interactions can be converted into user-item-explanation interactions, so as to build a connection between explanations and users/items. In the figure, $r_\ast$ represents a review, and $e_\ast$ represents an explanation sentence in the review.}
	\label{fig:interaction}
	\vspace{-15pt}
\end{figure}

Thus, we create three EXplanaTion RAnking datasets (denoted as EXTRA) for explainable recommendation research.
Specifically, they are built upon user generated reviews, which are the collection of users' real feedback towards items, and thus are an ideal proxy of explanation.
Also, the datasets can be further enriched by new explanations when the newly posted reviews contain new item features or up-to-date expressions.

However, simply adopting reviews \cite{WWW18-NARRE, EMNLP19-R2HP} or their sentences \cite{ICDM18-RL, AAAI19-DER} as explanations is less appropriate, because it is very unlikely for two reviews/sentences to be exactly the same, as a result, each review/sentence only appears once in the data.
For this reason, almost no user shares the same review/sentence (see r1 to r6 in Fig. \ref{fig:interaction}), which makes it difficult to design collaborative learning algorithms for explanation ranking.
Besides, not all sentences in a user review are of explanation purpose.
Our solution is to extract explanation sentences from reviews that co-occur across various reviews, so as to connect different user-item pairs with one particular explanation (e.g., u2-i1 and u3-i3 with explanation e2 in Fig. \ref{fig:interaction}), and build the user-item-explanation triplet interactions.
By finding the commonly used sentences, the quality of explanations such as readability and expressiveness can be guaranteed.
This type of textual explanations could be very effective in helping users make better and informed decisions.
A recent online experiment conducted on Microsoft Office 365 \cite{WWW20-Office} finds that their manually designed textual explanations, e.g., ``\textit{Jack shared this file with you}'', can help users to access documents faster.
It motivates us to automatically create this type of explanations for other application domains, e.g., movies, restaurants and hotels.

\begin{figure}
	\centering
	\includegraphics[scale=0.5]{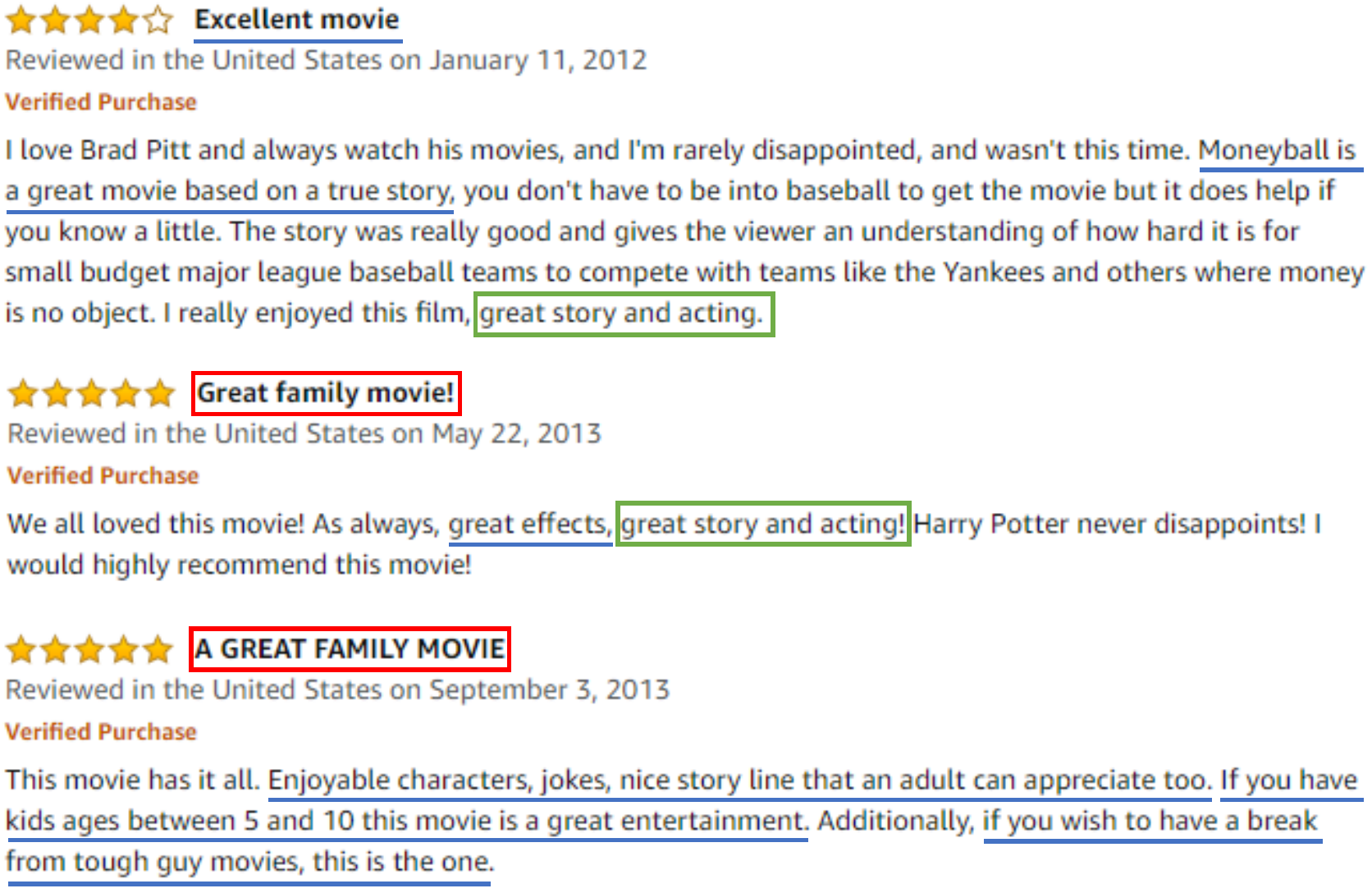}
	\caption{Three user reviews for different movies from Amazon (Movies \& TV category). Sentences of explanation purposes are highlighted in colors. Co-occurring explanations across different reviews are highlighted in rectangles.}
	\label{fig:example}
	\vspace{-10pt}
\end{figure}

Then, a follow-up problem is how to detect the nearly identical sentences across the reviews in a dataset.
Data clustering is infeasible to this case, because its number of centroids is pre-defined and fixed.
Computing the similarity between any two sentences in a dataset is practical but less efficient, since it has a quadratic time complexity.
To make this process more efficient, we develop a method that can categorize sentences into different groups, based on Locality Sensitive Hashing (LSH) \cite{Book11-LSH} which is devised for near-duplicates detection.
Furthermore, because some sentences are less suitable for explanation purpose (see the first review's first sentence in Fig. \ref{fig:example}), we only keep those sentences that contain both noun(s) and adjective(s), but not personal pronouns, e.g., ``\textit{I}''.
In this way, we can obtain high-quality explanations that talk about item features with certain opinions but do not go through personal experiences.
After the whole process, the explanation sentences remain personalized, since they resemble the case of traditional recommendation, where users of similar preferences write nearly identical review sentences, while similar items can be explained by the same explanations (see sentences in rectangles in Fig. \ref{fig:example}).

Notice that, our datasets are different from user-item-tag data \cite{PKDD07-FolkRank, WSDM10-PITF}, since a single tag when used as an explanation may not be able to clearly explain an item's specialty, e.g., a single word ``\textit{food}'' cannot well describe how good a restaurant's food tastes.

To sum up, our contributions are listed below:
\begin{itemize}
	\item We construct three large datasets consisting of user-item-explanation interactions, on which explainability can be evaluated via standard ranking metrics, e.g., NDCG.
	Datasets and code are made available online.\footnote{\url{https://github.com/lileipisces/EXTRA}}
	\item We address two key problems when creating such datasets, including the interactions between explanations and users/items, as well as the efficiency for grouping nearly identical sentences.
\end{itemize}

In the following, we first introduce our data processing approach and the resulting datasets in Section \ref{sec:method}.
Then, we present two explanation ranking formulations in Section \ref{sec:formulation}.
We experiment existing methods on the datasets in Section \ref{sec:experiment}.
Section \ref{sec:conclude} concludes this work.

\begin{figure*}
	\centering
	\subfigure[Naive way]{\includegraphics[scale=0.3]{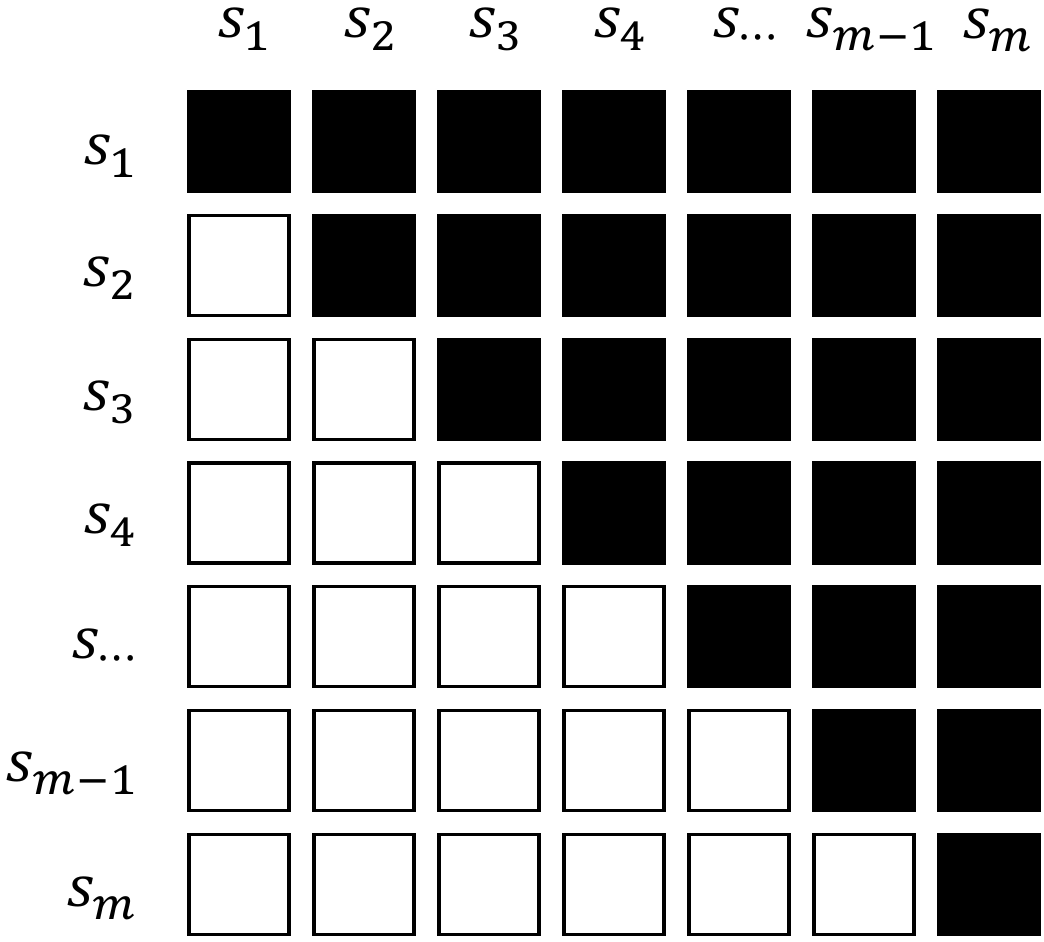}}
	\hspace{5mm}
	\subfigure[Sentence grouping: step 1]{\includegraphics[scale=0.3]{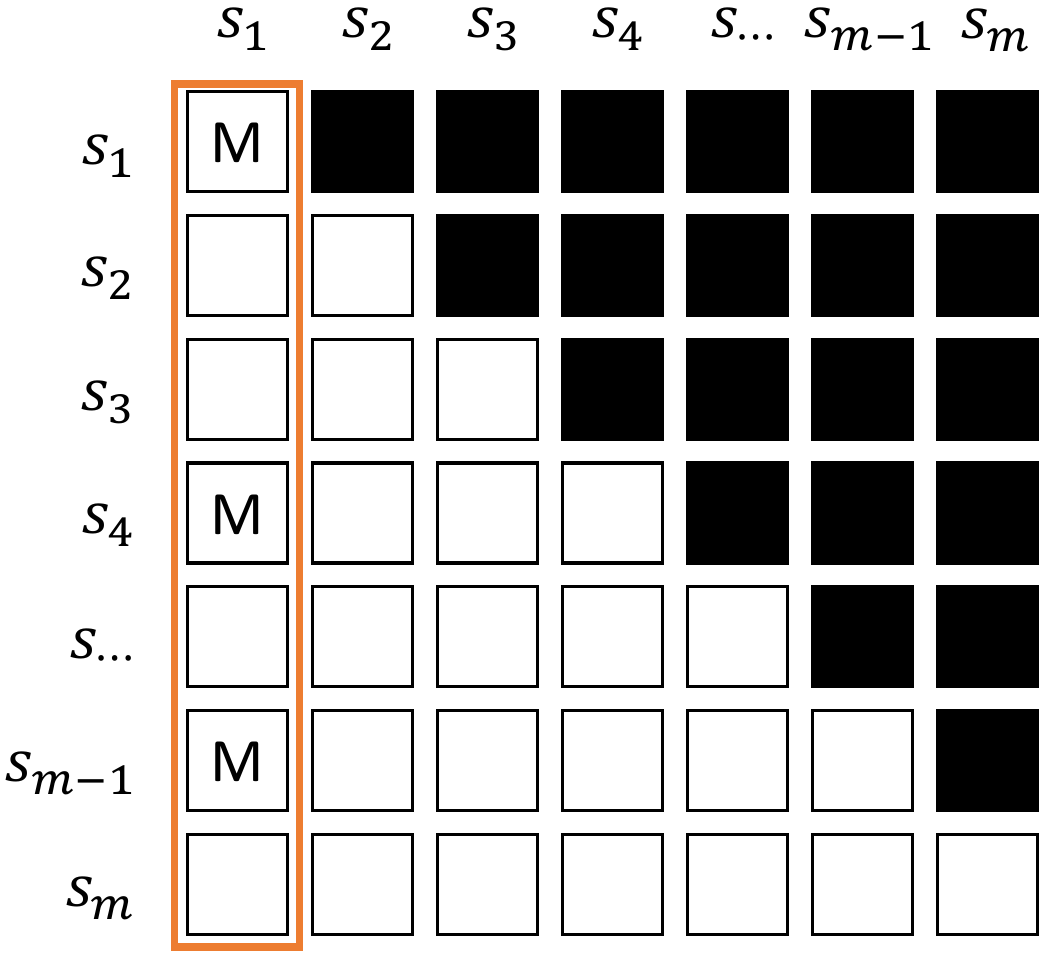}}
	\hspace{2mm}
	\subfigure[Sentence grouping: step 2]{\includegraphics[scale=0.3]{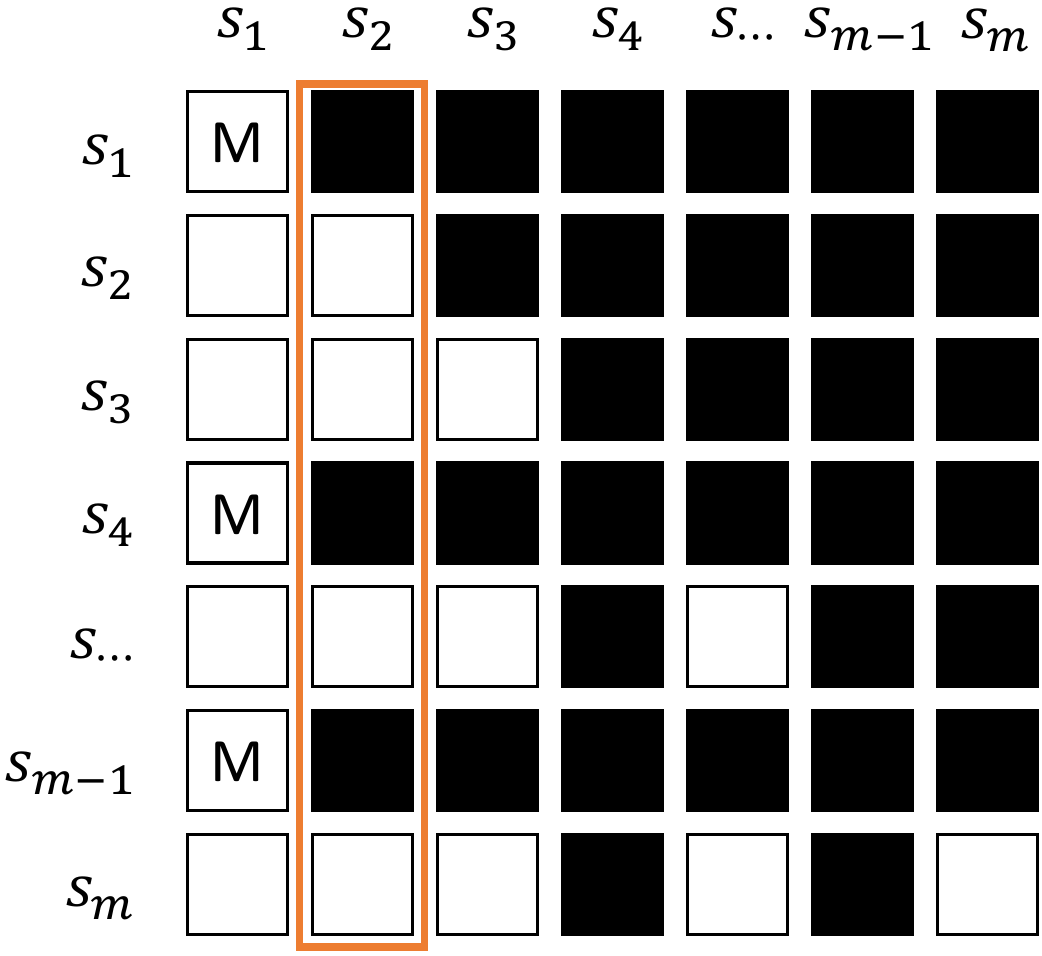}}
	\hspace{2mm}
	\subfigure[Sentence grouping: step 3]{\includegraphics[scale=0.3]{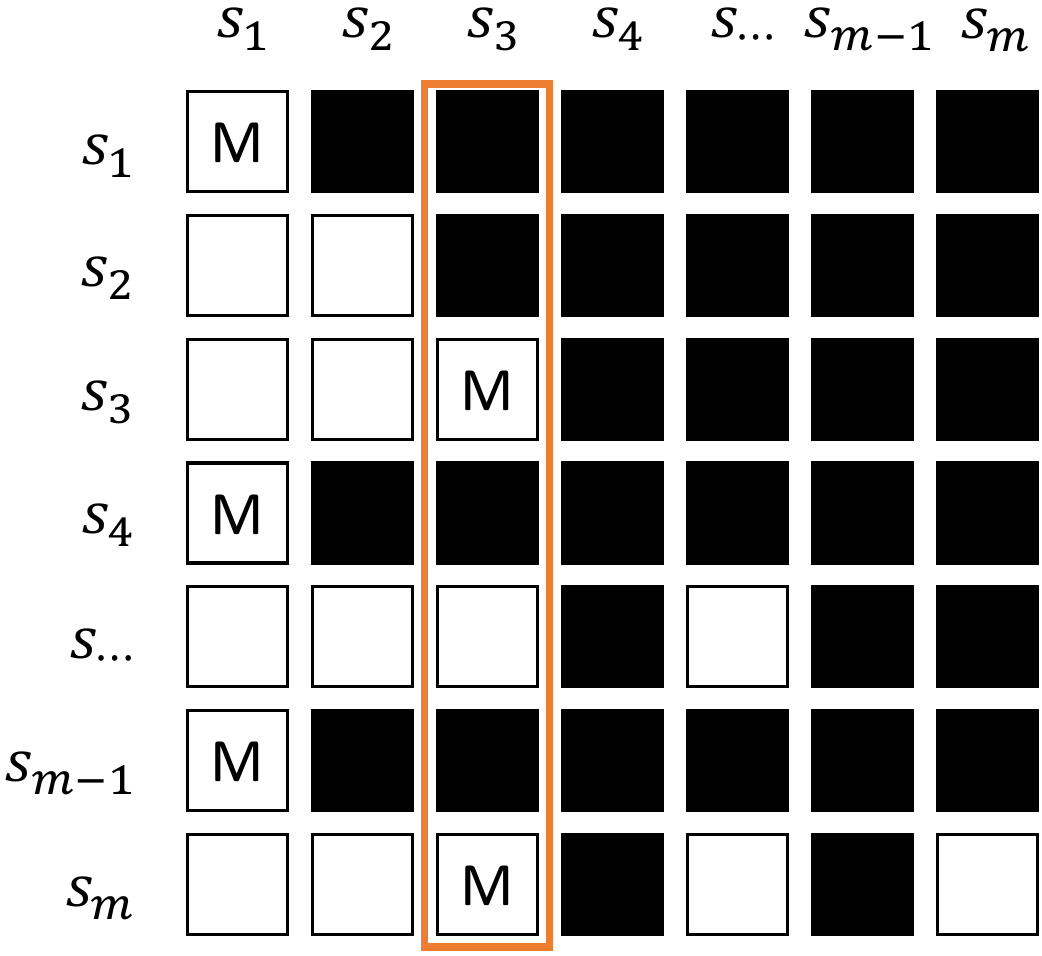}}
	\hspace{2mm}
	\subfigure[Sentence grouping: step 4]{\includegraphics[scale=0.3]{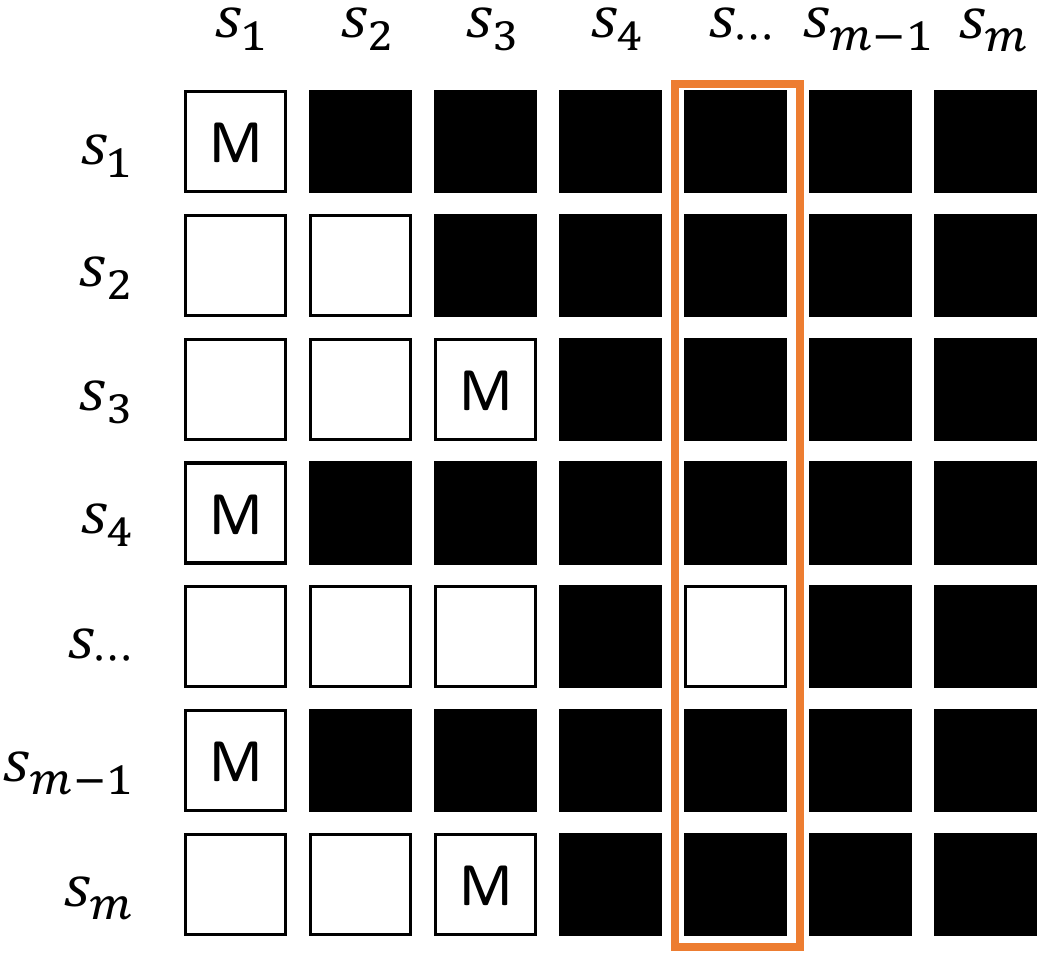}}
	\caption{White cells denote similarity computation, while black cells omit the computation.	(a) shows a naive way to compute the similarity between any two sentences, which would take quadratic time. (b)-(e) show four example steps in our more efficient sentence grouping algorithm, where orange rectangles denote query steps in LSH, and M denotes the matched duplicates.}
	\label{fig:process}
\end{figure*}

\begin{algorithm}
	\caption{Sentence Grouping via LSH}
	\label{alg:lsh}
	\begin{algorithmic}[1]
		\REQUIRE shingle size $n$, similarity threshold $t$, min group size $g$
		\ENSURE explanation set $\mathcal{E}$, groups of sentences $\mathcal{M}$
		\STATE Pre-process textual data to obtain the sentence collection $\mathcal{S}$
		\STATE $lsh \gets MinHashLSH(t)$, $\mathcal{C} \gets \varnothing$
		\FOR{sentence $s$ in $\mathcal{S}$}
		\STATE $m \gets MinHash()$  // create MinHash for $s$
		\FOR{$n$-shingle $h$ in $s$}
		\STATE $m.update(h)$  // convert $s$ into $m$ by encoding its $n$-shingles
		\ENDFOR
		\STATE $lsh.insert(m)$, $\mathcal{C}.add(m)$  // $\mathcal{C}$: set of all sentences' MinHash
		\ENDFOR
		\STATE $\mathcal{M} \gets \varnothing$, $\mathcal{Q} \gets \varnothing$  // $\mathcal{Q}$: set of queried sentences
		\FOR{$m$ in $\mathcal{C}$}
		\IF{$m$ not in $\mathcal{Q}$}
		\STATE $\mathcal{G} \gets lsh.query(m)$  // $\mathcal{G}$: ID set of duplicate sentences
		\IF{$\mathcal{G}.size > g$}
		\STATE $\mathcal{M}.add(\mathcal{G})$  // only keep groups with enough sentences
		\STATE $\mathcal{E}.add(\mathcal{G}.get())$  // keep one explanation in each group
		\ENDIF
		\FOR{$m'$ in $\mathcal{G}$}
		\STATE $lsh.remove(m')$, $\mathcal{Q}.add(m')$  // for efficiency
		\ENDFOR
		\ENDIF
		\ENDFOR
	\end{algorithmic}
\end{algorithm}

\section{Methodology and Results} \label{sec:method}

For explanation ranking, the datasets are expected to contain user-item-explanation interactions.
In this paper, we narrow down to the explanation sentences from user reviews.
The key problem is how to efficiently detect near-duplicates across different reviews, since it takes quadratic time to compute the similarity between any two sentences in a dataset.
In the following, we first present our approach to finding duplicate sentences based on sentence grouping, then introduce the data construction details, and at last analyze the datasets.

\subsection{Sentence Grouping}

The advantage of sentence grouping is three-fold.
First, it ensures the readability and expressiveness of the explanations, as they are extracted from real users' reviews based on the wisdom of the crowd.
Second, it allows the explanations to be connected with both users and items, so that we can design collaborative filtering models to learn and predict such connections.
Third, it makes explanation ranking and the automatic benchmark evaluation possible, since there are only a limited set of candidate explanations.

Computing the similarity between any two sentences in a dataset is computationally expensive. However, at each step of sentence grouping, it is actually unnecessary to compute the similarity for the already grouped sentences.
Therefore, we can reduce the computation cost by removing those sentences (see Fig. \ref{fig:process} (b)-(e) for illustration).
To find similar sentences more efficiently, we make use of Locality Sensitive Hashing (LSH) \cite{Book11-LSH}, which is able to conduct near-duplicate detection in sub-linear time.
LSH consists of three major steps.
First, a document (i.e., a sentence in our case) is converted into a set of $n$-shingles (a.k.a., $n$-grams).
Second, the sets w.r.t. all documents are converted to short signatures via hashing, so as to reduce computation cost and meanwhile preserve document similarity.
Third, the documents, whose similarity to a query document is greater than a pre-defined threshold, are returned.
The detailed procedure of sentence grouping is shown in Algorithm \ref{alg:lsh}.

Next, we discuss the implementation details.
To make better use of all the available text in a dataset, for each record we concatenate the review text and the heading/tip.
Then each piece of text is tokenized into sentences.
In particular, a sentence is removed if it contains personal pronouns, e.g., ``\textit{I}'' and ``\textit{me}'', since explanations are expected to be more objective than subjective.
We also calculate the frequency of nouns and adjectives in each sentence via NLTK\footnote{https://www.nltk.org}, and only keep the sentences that contain both noun(s) and adjective(s), so as to obtain more informative explanations that evaluate certain item features.
After the data pre-processing, we conduct sentence grouping via an open-source LSH \cite{Book11-LSH} package Datasketch\footnote{http://ekzhu.com/datasketch/lsh.html}.
Notice that, we apply it to all sentences in a dataset, rather than that of a particular item, because it is easier to find common expressions from a large amount of sentences.
When creating MinHash for each sentence, we set the shingle size $n$ to 2 so as to preserve the word ordering and meanwhile distinguish positive sentiment from negative sentiment (e.g., ``\textit{is good}'' v.s. ``\textit{not good}'').
We test the similarity threshold $t$ of querying sentences from [0.5, 0.6, ..., 0.9], and find that the results with 0.9 are the best.

\begin{table*}
	\centering
	\caption{Data format of our datasets. Each dataset contains two plain text files: IDs.txt and id2exp.txt. Entries in each line of the files are separated by double colon. In IDs.txt, expID denotes the explanation ID, which corresponds to a group of near-duplicate sentences, while the senID is the original sentence ID. When a record has multiple explanation IDs, they are separated by single colon. In id2exp.txt, expID applies to both expID and senID in IDs.txt.}
	\begin{tabular}{l|l}
		\hline
		File & Format \\ \hline
		\multirow{3}{*}{IDs.txt} & \textbf{userID::itemID::rating::timeStamp::expID:expID::senID:senID} \\ \cline{2-2}
		& A20YXFTS3GUGON::B00ICWO0ZY::5::1405958400::13459471:5898244::32215058:32215057 \\ \cline{2-2}
		& APBZTFB6Y3TUX::B000K7VHPU::5::1394294400::13459471::21311508 \\ \hline
		
		\multirow{4}{*}{id2exp.txt} & \textbf{expID::expSentence} \\ \cline{2-2}
		& 5898244::Great Movie \\ \cline{2-2}
		& 13459471::This is a wonderful movie \\ \cline{2-2}
		& 21311508::This is a wonderful movie \\ \hline
	\end{tabular}
	\label{tbl:format}
\end{table*}

\subsection{Data Construction}

We construct our datasets on three domains: Amazon Movies \& TV\footnote{http://jmcauley.ucsd.edu/data/amazon} (movie), TripAdvisor\footnote{https://www.tripadvisor.com} (hotel) and Yelp\footnote{https://www.yelp.com/dataset/challenge} (restaurant).
In each of the datasets, a record is comprised of user ID, item ID, overall rating in the scale of 1 to 5, and textual review.
After splitting reviews into sentences, we apply sentence grouping (in Algorithm \ref{alg:lsh}) over them to obtain a large amount of sentence groups.
A group is removed if its number of sentences is smaller than 5 so as to retain commonly seen explanations.
We then assign each of the remaining groups an ID that we call an explanation ID.
Eventually, a user-item pair may be connected to none, one or multiple explanation IDs since the review of the user-item pair may contain none, one or multiple explanation sentences.
We remove the user-item pairs that are not connected to any explanation ID, and the remaining records are thus user-item-explanation triplets.

To make our datasets more friendly to the community, we largely follow the data format of a well-known dataset MovieLens\footnote{https://grouplens.org/datasets/movielens/}.
Specifically, we store each processed dataset in two separate plain text files: \textbf{IDs.txt} and \textbf{id2exp.txt}.
The former contains the meta-data information, such as user ID, item ID and explanation ID, while the latter stores the textual content of an explanation that can be retrieved via the explanation ID.
The entries of each line in both files are separated by double colon, i.e., ``::''.
If a line in IDs.txt contains multiple explanation IDs, they are separated by a single colon, i.e., ``:''.
The detailed examples are shown in Table \ref{tbl:format}.
With this type of data format, loading the data would be quite easy, but we also provide a script in our code for data loading.

\subsection{Data Analysis}

Table \ref{tbl:dataset} shows the statistics of the processed datasets.
Notice that, multiple explanations may be detected in a review, which leads to more than one user-item-explanation triplets.
As we can see, all the three datasets are very sparse.

\begin{table}
	\centering
	\caption{Statistics of the datasets. Density is the \#triplets divided by \#users $\times$ \#items $\times$ \#explanations.}
	\setlength{\tabcolsep}{3.5pt}
	\begin{tabular}{l|r|r|r}
		\hline
		& Amazon & TripAdvisor & Yelp \\
		\hline
		\# of users                & 109,121                & 123,374               & 895,729 \\
		\# of items                 & 47,113                & 200,475               & 164,779 \\
		\# of explanations               & 33,767              & 76,293            & 126,696 \\
		\# of $(u, i)$ pairs & 569,838 & 1,377,605 & 2,608,860 \\
		\# of $(u, i, e)$ triplets    & 793,481                & 2,618,340               & 3,875,118 \\
		\# of explanations / $(u, i)$ pair & 1.39 & 1.90 & 1.49 \\
		Density ($\times 10^{-10}$)  & 		45.71   	&  	13.88			  & 2.07 \\
		\hline
	\end{tabular}
	\label{tbl:dataset}
\end{table}

Next, we show 5 example explanations on each dataset in Table \ref{tbl:case}.
We can see that the explanations vary from dataset to dataset, but they all reflect the characteristics of the corresponding datasets, e.g., ``\textit{a wonderful movie for all ages}'' on the dataset Amazon Movies \& TV.
The occurrence of short explanations is high, not only because LSH favors short text, but also because people tend to express their opinions using common and concise phrases.
Moreover, we can observe some negative expressions, which can be used to explain dis-recommendations \cite{SIGIR14-EFM}.

Because constructing the datasets does not involve manual efforts, we do observe one minor issue.
Since a noun is not necessarily an item feature, the datasets contain a few less meaningful explanations that are less relevant to items, e.g., ``\textit{the first time}''.
This issue can be effectively addressed if we pre-define a set of item features or filter out item-irrelevant nouns for each dataset.

\begin{table}
	\centering
	\caption{Example explanations after sentence grouping on three datasets. Occurrence means the number of near duplicate explanations.}
	\begin{tabular}{l|l}
		\hline
		\textbf{Explanation} & \textbf{Occurrence} \\ \hline
		\hline
		\multicolumn{2}{c}{\textbf{Amazon Movies \& TV}} \\ \hline
		Excellent movie & 3628 \\ \hline
		This is a great movie & 2941 \\ \hline
		Don't waste your money & 834 \\ \hline
		The sound is okay & 11 \\ \hline
		A wonderful movie for all ages & 6 \\ \hline
		\hline
		\multicolumn{2}{c}{\textbf{TripAdvisor}} \\ \hline
		Great location & 61993 \\ \hline
		The room was clean & 6622 \\ \hline
		The staff were friendly and helpful & 2184 \\ \hline
		Bad service & 670 \\ \hline
		Comfortable hotel with good facilities & 8 \\ \hline
		\hline
		\multicolumn{2}{c}{\textbf{Yelp}} \\ \hline
		Great service & 46413 \\ \hline
		Everything was delicious & 5237 \\ \hline
		Prices are reasonable & 2914 \\ \hline
		This place is awful & 970 \\ \hline
		The place was clean and the food was good & 6 \\ \hline \hline
	\end{tabular}
	\label{tbl:case}
\end{table}

\begin{table*}
	\centering
	\caption{Performance comparison of all methods on the top-10 explanation ranking in terms of NDCG, Precision (Pre), Recall (Rec) and F1 (\%). The best performing values are boldfaced.}
	\setlength{\tabcolsep}{3.3pt}
	\begin{tabular}{l|cccc|cccc|cccc}
		\hline
		\multicolumn{1}{c|}{\multirow{2}{*}{}} & \multicolumn{4}{c|}{Amazon} & \multicolumn{4}{c|}{TripAdvisor} & \multicolumn{4}{c}{Yelp} \\ \cline{2-13}
		\multicolumn{1}{c|}{}  & NDCG@10 & Pre@10 & Rec@10 & F1@10 & NDCG@10 & Pre@10 & Rec@10 & F1@10 & NDCG@10 & Pre@10 & Rec@10 & F1@10 \\ \hline
		CD & 0.001 & 0.001 & 0.007 & 0.002 & 0.001 & 0.001 & 0.003 & 0.001 & 0.000 & 0.000 & 0.003 & 0.001 \\
		RAND & 0.004 & 0.004 & 0.027 & 0.006 & 0.002 & 0.002 & 0.011 & 0.004 & 0.001 & 0.001 & 0.007 & 0.002 \\
		RUCF & 0.341 & 0.170 & 1.455 & 0.301 & 0.260 & 0.151 & 0.779 & 0.242 & 0.040 & 0.020 & 0.125 & 0.033 \\
		RICF & 0.417 & 0.259 & 1.797 & 0.433 & 0.031 & 0.020 & 0.087 & 0.030 & 0.037 & 0.026 & 0.137 & 0.042 \\
		PITF & \textbf{2.352} & \textbf{1.824} & \textbf{14.125} & \textbf{3.149} & \textbf{1.239} & \textbf{1.111} & \textbf{5.851} & \textbf{1.788} & \textbf{0.712} & \textbf{0.635} & \textbf{4.172} & \textbf{1.068} \\ \hline
	\end{tabular}
	\label{tbl:exp}
\end{table*}

\section{Explanation Ranking Formulation} \label{sec:formulation}

The task of explanation ranking aims at finding a list of explanations to explain a recommendation for a user.
Similar to item ranking, these explanations are better to be personalized to the user's interests as well as the target item's characteristics.
To produce such a personalized explanation list, a recommender system can leverage the user's historical data, e.g., her past interactions and comments on other items.
In the following, we introduce two types of explanation ranking formulation, including global-level and item-level explanation ranking.

In the setting of \textbf{global-level explanation ranking}, there is a collection of explanations $\mathcal{E}$ that are globally shared for all items.
The recommender system can estimate a score $\hat{r}_{u, i, e}$ for each explanation $e \in \mathcal{E}$ for a given pair of user $u \in \mathcal{U}$ and item $i \in \mathcal{I}$.
The user-item pair can be either an item that a user interacted before, or an item recommended for the user based on a recommendation model.
According to the scores, the top-$N$ explanations can be selected to justify why recommendation $i$ is made for user $u$.
Formally, this explanation list can be defined as:
\begin{equation}
	\text{Top}(u, i, N) := \mathop{\arg\max}_{e \in \mathcal{E}}^{N} \hat{r}_{u, i, e}
	\label{eqn:topn}
\end{equation}

Meanwhile, we can perform \textbf{item-level explanation ranking} to select explanations from the target item's collection, which can be formulated as:
\begin{equation}
	\text{Top}(u, i, N) := \mathop{\arg\max}_{e \in \mathcal{E}_i}^{N} \hat{r}_{u, i, e}
	\label{eqn:topni}
\end{equation}
where $\mathcal{E}_i$ is item $i$'s explanation collection.

The two formulations respectively have their own advantages.
The global-level ranking can make better use of all the user-item-explanation interactions, e.g., ``\textit{great story and acting}'' for different items (see Fig. \ref{fig:example}), so as to better capture the relation between users, items and explanations.
As a comparison, item-level ranking can prevent the explanation model from presenting item-dependent explanations to irrelevant items,
e.g., ``\textit{Moneyball is a great movie based on a true story}'' that only applies to the movie \textit{Moneyball}.
Depending on the application scenarios, we may adopt different formulations.

\section{Experiments} \label{sec:experiment}

In this section, we first introduce five prototype methods for explanation ranking.
Then, we discuss the experimental details.
At last, we analyze the results of different methods.

\subsection{Explanation Ranking Methods}

On the global-level explanation ranking task, we test five methods.
The first one is denoted as RAND, which randomly selects explanations from the explanation set $\mathcal{E}$ for any given user-item pair.
It is simply used to show the bottom line performance of explanation ranking.
The other four methods can be grouped into two categories, including collaborative filtering and tensor factorization.
For the ranking purpose, each of the four methods must estimate a score $\hat{r}_{u, i, e}$ for a triplet $(u, i, e)$.

\subsubsection{Collaborative Filtering}

Collaborative Filtering (CF) \cite{CSCW94-UCF, WWW01-ICF} is a typical type of recommendation algorithms that recommend items for a user, based on either the user's neighbors who have similar preference, or each item's neighbors.
It naturally fits the explanation ranking task, as some users may care about certain item features, and some items' specialty could be similar.
We extend user-based CF (UCF) and item-based CF (ICF) to our ternary data, following \cite{PKDD07-FolkRank}, and denote them as RUCF and RICF, where ``R'' means ``Revised''.
Taking RUCF as an example, we first compute the similarity between two users $u$ and $u^\prime$ via Jaccard Index as follows,
\begin{equation}
	s_{u, u'} = \frac{\vert \mathcal{E}_u \cap \mathcal{E}_{u'} \vert}{\vert \mathcal{E}_u \cup \mathcal{E}_{u'} \vert}
\end{equation}
where $\mathcal{E}_u$ and $\mathcal{E}_{u'}$ denote the explanations associated with $u$ and $u'$, respectively.
Then we estimate a score for the triplet $(u, i, e)$, for which we only retain user $u$'s neighbors who interacted with both item $i$ and explanation $e$.
\begin{equation}
	\hat{r}_{u, i, e} = \sum_{u' \in \mathcal{N}_u \cap (\mathcal{U}_i \cap \mathcal{U}_e)} s_{u, u'}
	\label{eqn:icf}
\end{equation}
Similarly, RICF can predict a score for the same triplet via the neighbors of items.

\subsubsection{Tensor Factorization}

The triplets formed by users, items and explanations correspond to entries in an interaction cube, whose missing values could be recovered by Tensor Factorization (TF) methods.
Thus, we test two typical TF methods, including Canonical Decomposition (CD) \cite{Springer70-CD} and Pairwise Interaction Tensor Factorization (PITF) \cite{WSDM10-PITF}.
To predict a score $\hat{r}_{u, i, e}$, CD performs element-wise multiplication on the latent factors of user $u$, item $i$ and explanation $e$, and then sums over the resultant vector.
Formally, it can be written as:
\begin{equation}
\hat{r}_{u, i, e} = (\mathbf{p}_u \odot \mathbf{q}_i)^\top \mathbf{o}_e = \sum_{k = 1}^d p_{u, k} \cdot q_{i, k} \cdot o_{e, k}
\label{eqn:cd}
\end{equation}
where $\odot$ represents two vectors' element-wise multiplication, and $d$ is the number of latent factors.
PITF does the prediction via two sets of matrix multiplication as follows,
\begin{equation}
	\hat{r}_{u, i, e} = \mathbf{p}_u^\top \mathbf{o}_e^U + \mathbf{q}_i^\top \mathbf{o}_e^I = \sum_{k = 1}^d p_{u, k} \cdot o_{e, k}^U + \sum_{k = 1}^d q_{i, k} \cdot o_{e, k}^I
	\label{eqn:pitf}
\end{equation}
where $\mathbf{o}_e^U$ and $\mathbf{o}_e^I$ are two different latent factors for the same explanation.

We opt for Bayesian Personalized Ranking (BPR) criterion \cite{UAI09-BPR} to learn the parameters of the two TF methods, because it can model the relative ordering of explanations, e.g., the ranking score of a user's interacted explanations should be greater than that of her un-interacted explanations.
The objective function of both CD and PITF is shown below:
\begin{equation}
	\min_{\Theta} \sum_{u \in \mathcal{U}} \sum_{i \in \mathcal{I}_u} \sum_{e \in \mathcal{E}_{u, i}} \sum_{e' \in \mathcal{E} / \mathcal{E}_{u, i}} - \ln \sigma (\hat{r}_{u, i, ee'}) + \lambda \left| \left| \Theta \right| \right|_F^2
	\label{eqn:exp}
\end{equation}
where $\hat{r}_{u, i, ee'} = \hat{r}_{u, i, e} - \hat{r}_{u, i, e'}$ denotes the difference between two interactions, $\sigma(\cdot)$ is the sigmoid function, $\mathcal{I}_{u}$ represents user $u$'s interacted items, $\mathcal{E}_{u, i}$ is the explanation set of $(u, i)$ pair for training, $\Theta$ denotes model parameters, and $\lambda$ is a coefficient for preventing the model from over-fitting.
To learn model parameters $\Theta$, we optimize Eq. \eqref{eqn:exp} for both CD and PITF via stochastic gradient descent.
At the testing stage, we can measure scores of explanations in $\mathcal{E}$ for a user-item pair, and then rank them according to Eq. \eqref{eqn:topn}.

Notice that, CD and PITF may be further enriched by considering more complex relationships between explanations (e.g., the ranking score of a user's positive explanations $>$ the other users' explanations $>$ the user's negative explanations).
We leave the exploration for future work.

\subsection{Experimental Settings}

As discussed earlier, this paper aims at achieving a standard way of evaluating recommendation explanations via ranking.
To compare the performance of different methods on the explanation ranking task, we adopt four ranking-oriented metrics: Normalized Discounted Cumulative Gain (\textbf{NDCG}), Precision (\textbf{Pre}), Recall (\textbf{Rec}) and \textbf{F1}.
Top-10 explanations are returned for each testing user-item pair.
We randomly select 70\% of the triplets in each dataset for training, and the rest for testing.
Also, we make sure that the training set holds at least one triplet for each user, item and explanation.
We do this for 5 times, and thus obtain 5 data splits, on which we report the average performance of each method.

All the methods are implemented in Python.
To allow CF-based methods (i.e., RUCF and RICF) better utilize user/item neighbors, we do not restrict the upper limit of size for $\mathcal{N}_u$ and $\mathcal{N}_i$.
For TF-based methods, i.e., CD and PITF, we search the number of latent factors $d$ from [10, 20, 30, 40, 50], regularization coefficient $\lambda$ from [0.001, 0.01, 0.1], learning rate $\gamma$ from [0.001, 0.01, 0.1], and maximum iteration number $T$ from [100, 500, 1000].
After parameter tuning, we use $d = 20$, $\lambda = 0.01$, $\gamma = 0.01$ and $T = 500$ for both CD and PITF.

\begin{table}
	\centering
	\caption{Top-5 and bottom-5 explanations ranked by PITF for a user-item pair on Amazon Movies \& TV dataset. There are two ground-truth explanations, and the matched one is boldfaced.}
	\begin{tabular}{l|l}
		\hline \hline
		\textbf{Top-5} & \textbf{Bottom-5} \\ \hline \hline
		The acting is superb & Good B-movie \\ \hline
		\textbf{The cast is first rate} & The final Friday \\ \hline
		The acting is wonderful & This was a great event \\ \hline
		The main character & Dead alive \\ \hline
		The acting is first rate & A voice teacher and early music fan \\ \hline
	\end{tabular}
	\label{tbl:rankcase}
\end{table}

\subsection{Results and Analysis}

Table \ref{tbl:exp} presents the performance comparison of different methods on three datasets.
We have the following observations.
First, each method performs consistently on the three datasets regarding the four metrics.
Second, the performances of both RAND and CD are the worst, because RAND is non-personalized, while the data sparsity problem (see Table \ref{tbl:dataset}) may be difficult to mitigate for CD that simply multiplies three latent factors.
Third, both RUCF and RICF that can make use of user/item neighbors, are better than RAND, but they are still limited because of the data sparsity issue.
Lastly, PITF improves CD and also outperforms RUCF and RICF, with its specially designed model structure to tackle data sparsity issue (see \cite{WSDM10-PITF} for discussion).

To better understand how explanation ranking works, in Table \ref{tbl:rankcase} we provide top-5 and bottom-5 explanations ranked by PITF for a specific user-item pair.
For this record, there are two ground-truth explanations, i.e., ``\textit{The story is true}'' and ``\textit{The cast is first rate}'', and the latter is ranked the second by PITF.
Moreover, the key features of the other explanations among the top-5 are ``\textit{character}'' and ``\textit{acting}'', which are semantically close to ``\textit{cast}''.
As a comparison, the bottom-5 explanations are less relevant to the ground-truth, and their sentence quality is not good.
This hence shows the effectiveness of explanation ranking in finding relevant and high-quality explanations for recommendations.

\section{Conclusion and Future Work} \label{sec:conclude}

In this paper, we construct three explanation ranking datasets for explainable recommendation research, with an attempt to achieve a standard way of evaluating explainability.
To this end, we address two problems during data construction, including the lack of user-item-explanation interactions and the efficiency of detecting similar sentences from user reviews.
Since this paper's focus is about data construction, we present our explanation ranking methods (with and without utilizing the textual content of explanations) in \cite{21-L2E}.

We believe the explanation task can be formalized as a standard ranking task just as the recommendation task. This not only enables standard evaluation of explainable recommendation but also helps to develop advanced explainable recommendation models. In the future, we will extend this work on two dimensions. One dimension is to develop multimodal explanation ranking datasets by adopting our sentence grouping approach to images, so as to construct datasets with visual explanations. Another dimension is to develop better explanation ranking models for explainable recommendation \cite{FTIR20-Survey}. Moreover, we intend to seek industrial cooperation for conducting online experiments to test the impact of the ranked explanations to real users, e.g., the click through rate, which will help to validate explainable recommendation models under various kinds of recommendation scenarios.

\begin{acks}
This work was partially supported by Hong Kong Research Grants Council (RGC) (project RGC/HKBU12201620) and partially supported by NSF IIS-1910154 and IIS-2007907.
Any opinions, findings, conclusions or recommendations expressed in this material are those of the authors and do not necessarily reflect those of the sponsors.
\end{acks}

\bibliographystyle{ACM-Reference-Format}
\balance
\bibliography{bibliography}

\appendix

\end{document}